\documentclass[apjl]{emulateapj}

\usepackage{natbib}
\slugcomment{Accepted in ApJ. COPYRIGHT 2010. All rights reserved.}

\shorttitle{VVC}
\shortauthors{Mawet et al.}

\begin{document}


\title{The Vector Vortex Coronagraph: Laboratory Results \\
and First Light at Palomar Observatory}


\author{D. Mawet\altaffilmark{1}, E. Serabyn, K. Liewer, R. Burruss}
\affil{Jet Propulsion Laboratory, California Institute of Technology, Pasadena, CA 91109, USA}

\author{J. Hickey}
\affil{Palomar Observatory, California Institute of Technology, P.O. Box 200, Palomar Mountain, CA 92060, USA}

\author{D. Shemo}
\affil{JDS Uniphase Corporation, 2789 Northpoint Parkway, Santa Rosa, CA 95407, USA, USA}


\email{Dimitri.Mawet@jpl.nasa.gov}


\altaffiltext{1}{Nasa Postdoctoral Fellow.}



\begin{abstract}
High-contrast coronagraphy will be needed to image and characterize faint extra-solar planetary systems. Coronagraphy is a rapidly evolving field, and many enhanced alternatives to the classical Lyot coronagraph have been proposed in the past ten years. Here, we discuss the operation of the vector vortex coronagraph, which is one of the most efficient possible coronagraphs. We first present recent laboratory results, and then first light observations at the Palomar observatory. Our near-infrared H-band (centered at $\sim 1.65 \mu$m) and K-band (centered at $\sim 2.2 \mu$m) vector vortex devices demonstrated excellent contrast results in the lab, down to $\sim 10^{-6}$ at an angular separation of $\sim 3 \lambda/d$. On sky, we detected a brown dwarf companion 3000 times fainter than its host star (HR~7672) in the K$_s$ band (centered at $\sim 2.15 \mu$m), at an angular separation of $\sim 2.5 \lambda/d$. Current and next-generation high-contrast instruments can directly benefit from the demonstrated capabilities of such a vector vortex: simplicity, small inner working angle, high optical throughput ($>90\%$), and maximal off-axis discovery space.
\end{abstract}





\section{Introduction}\label{sect:intro}

High contrast imaging of extrasolar planetary systems represents one of the most challenging technological goals of modern observational astrophysics. The challenge is twofold: the small angular separation between exoplanets and their host stars requires a high angular resolution (e.g.~100 mas for a 1 AU distance at 10 pc), and the brightness contrast between them is tremendous, ranging from $10^{-3}$ for very close hot giant planets, up to $10^{-10}$ for Earth-like planets. After the first resolved observation of a bound planetary mass object was obtained by \citet{Chauvin05a}, several exoplanets have recently been imaged. The majority of these observations were eased by the fairly large distance of the companions from their host stars ($0\farcs 5-15\arcsec$) and the moderate star/planet contrast in the near-infrared ($\Delta m \approx 5-12$) due to the relatively high thermal emission of the young forming planets (at $\sim$1000 K). The best characterizations obtained so far have relied on accurate multi-wavelength broadband photometry in the visible and near-infrared, and on multi-epoch astrometry. Keplerian motion was unambiguously detected for the most recent discoveries of \citet{Kalas08} around Fomalhaut and \citet{Marois08} around HR~8799. 

Direct imaging is clearly a self-consistent and complete characterization method as it provides a straightforward means to constrain orbital parameters and perform spectro-photometric measurements in the most varied orbital configurations, at almost any point in time. However, the few exoplanets and the majority of circumstellar disks imaged so far all required high contrast \citep{AbsilMawet10,Oppenheimer09,Duchene08, Wyatt08}. Image dynamic ranges can be increased with a coronagraph, and in a comparison of coronagraphs, \citet{Guyon2006} identified two as closest to theoretically ideal in terms of ``useful throughput'': the phase-induced amplitude apodization coronagraph, and the optical vortex phase-mask coronagraph \citep{Mawet2005a,Foo2005,Jenkins2008,Swartzlander2008}. The optical vortex is truly close to ideal, as it has a small inner working angle (IWA), high throughput, and a completely clear off-axis discovery space. We have therefore been pursuing an especially promising version of the vortex coronagraph, the vector vortex coronagraph (VVC). Here we present our most recent laboratory results with VVC masks based on birefringent liquid crystal polymers (LCPs). These tests were carried out in the near-infrared, as the masks considered here are intended for use behind ground-based extreme adaptive optics systems (ExAO). We also present the results and consider the implication of the first use of our VVC on the sky with the Palomar ExAO-level well-corrected subaperture \citep[WCS][]{Serabyn07,Serabyn09}.

\section{The Vector Vortex Coronagraph}\label{sect:ovc}
\begin{figure*}[!ht]
  \centering
\includegraphics[scale=0.44]{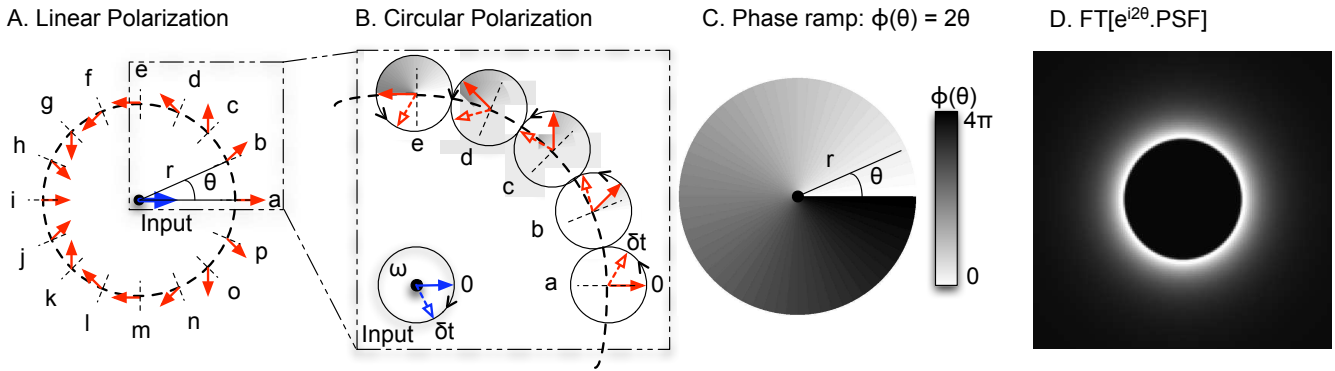}
  \caption{The VVC azimuthal phase ramp. Panel A: rotationally symmetric HWP with an optical axis orientation that rotates about the center (dashed lines perpendicular to the circumference). The net effect of a HWP on a linear impinging polarization is to rotate it by $-2\times \alpha$ where $\alpha$ is the angle between the incoming polarization direction and the fast optical axis. An incoming horizontal polarization (blue arrow) is transformed by the vector vortex so that it spins around its center twice as fast as the azimuthal coordinate $\theta$ (red arrows). Panel~B: for circular polarization, the output field rotation is strictly equivalent to a phase delay (the starting angle $0$ is rotated; therefore phase shifted). The angle of local rotation of the polarization vector corresponds to a ``geometrical'' phase: upon a complete rotation about the center of the rotationally symmetric HWP, it has undergone a total $2\times 2\pi$ phase ramp, which corresponds to the definition of an optical vortex of topological charge 2 (panel~C). Upon propagation from the focal plane to the subsequent pupil plane, the Fourier transform (FT) of the product of the PSF by the azimuthal phase ramp sends the light outside the original pupil area (Panel~D). \label{ovvc2pol}}
\end{figure*}

An optical vortex is generated by a phase screw of the form $e^{i\theta}$, with $\theta$ being the azimuthal coordinate. There are two kinds of optical vortex known: the scalar optical vortex, implemented by a structural helix \citep{Foo2005,Swartzlander2008} providing a scalar optical phase delay (OPD) which applies to the two orthogonal polarization components of natural light identically, and the vectorial vortex, implemented by a rotationally symmetric halfwave plate (HWP), providing a ``geometrical'' phase shift \citep{Pancha56, Berry87} that applies opposite phase screws to the two orthogonal circular polarization states (Fig.~\ref{ovvc2pol}).

The detailed theory of the vector vortex coronagraph is provided in \citet{Mawet2005a, Mawet2009}, and so here we merely provide a quick schematic overview. In the vector vortex, for a linearly polarized input field (or for natural light projected onto a linear basis), the rotationally symmetric HWP rotates the polarization vector as in Fig.~\ref{ovvc2pol}a. The definition of circular polarization is just a linear polarization rotating at the angular frequency $\omega$ (equal to the pulsation of the electromagnetic field), so that a rotation $\phi = 2\theta$ of the polarization vector is strictly equivalent to a phase delay (Fig.~\ref{ovvc2pol}b). If, at any given point in space, the polarization vector is rotated such as in Fig.~\ref{ovvc2pol}a, it implies for the circular polarization (Fig.~\ref{ovvc2pol}b), that it has acquired a geometric phase ramp $e^{i\phi}=e^{i2\theta}$ such as represented in Fig.~\ref{ovvc2pol}c. $\phi$ thus represents both an angle and a phase -- hence the term ``geometrical'' phase. 
 
The factor $2$ before the azimuthal coordinate in $e^{i2\theta}$ is called the ``topological charge'' $l$. It determines the polarization spin rate, and the subsequent height of the phase ramp after a full $2\pi$ rotation. In this particular example, we considered a vortex of topological charge 2, meaning that, upon a complete rotation about the center, it has undergone a total $2\times 2\pi$ phase shift. Note that this azimuthal phase ramp is generated without a corresponding structural helix. When the point-spread function (PSF) at the focal plane of a telescope is centered on such a vector vortex, for non-zero even values of $l$, the light, upon propagation to a subsequent pupil plane, then appears entirely outside the original pupil area, where it can be rejected by a Lyot stop the same size or slightly smaller than the entrance pupil \citep[Fig.~\ref{ovvc2pol}d,][]{Mawet2005a}.

\section{LCP technology and current devices}\label{sec:tech}

We have been developing rotationally symmetric HWPs based on LCPs, which  combine the birefringent properties of liquid crystals with the excellent mechanical properties of polymers. The orientation of a LCP is achieved through photo-alignment. Once aligned and cured, the polymer reaches a very stable solid state. The detailed proprietary fabrication process of our manufacturer JDSU has already been described in \citet{Mawet2009}. In that paper, we presented the theory of operation and obtained encouraging results, experimentally validating the VVC concept and showing good agreement with an empirical model of the imperfections. The first sample was mainly limited by the size (diameter $\sim 100$ $\mu$m) of the central so-called ``region of disorientation'', where the LCP loses the appropriate orientation close to the vortex axis, where most of the energy of the Airy diffraction pattern of the central stellar source would land. Our goal for the next generation of devices was thus to significantly decrease the central region of disorientation, and to cover the reduced residual zone by an opaque Aluminum mask so as to prevent leakage thru the center of the mask. Numerical simulations in \citet{Mawet2009} promised dramatic improvements in contrast if these steps were taken. The reduction was to be enabled by a redesign of the mechanical manufacturing apparatus described in \citet{Mawet2009}.

\begin{figure}[!hb]
  \centering
\includegraphics[scale=0.5]{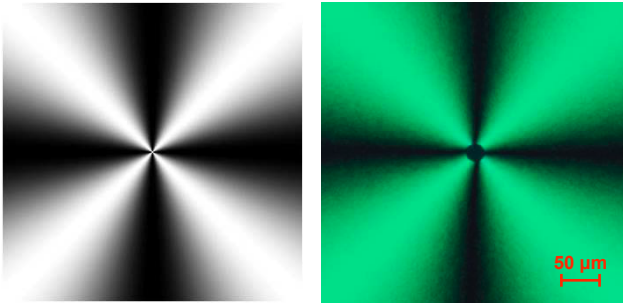}
  \caption{Left: theoretical pattern showing the calculated periodic modulation of the intensity transmitted by a VVC of topological charge 2 between crossed linear polarizers, corresponding to the polarization rotation shown in Fig.~\ref{ovvc2pol}a. Right: actual topological charge 2 K-band VVC device under a polarizing microscope (between crossed linear polarizers). Note the good centering of the Al spot (diameter 25 $\mu$m) on the vortex center and the almost perfect symmetry of the pattern, which is very close to the theoretical one (left). \label{ovvc2sample}}
\end{figure}

The new topological charge 2 H- and K-band devices achieved our manufacturing goals. The size of the central region of disorientation was reduced to a diameter of $\sim25$ $\mu$m, a 4-fold improvement upon the first vortex device. We chose to cover the remaining residual zone by an Aluminum spot 25 $\mu$m in diameter (Fig.~\ref{ovvc2sample}, right). Given any beam $F$ number $\geq 10$ ($F=f/d$, with $f$ the focal length, and $d$ the diameter of the beam as defined by the entrance pupil), the radius of the spot is comparable to or smaller than the beam width $F\lambda$ (where $\lambda$ is the wavelength of the incident light), so the Al spot does not affect the IWA at all. Note that classical and apodized Lyot coronagraphs use much larger central opaque spots, many $F\lambda$ across.

The polarization properties of the new devices were measured on an Axoscan Mueller Matrix Spectro-Polarimeter. These measurements allowed us to calculate the central wavelengths (where the retardance is $\pi$ radians) of the H and K devices to be 1642.5 nm and 2225.2 nm, respectively, reasonably well centered on the H and K bands. Given the birefringence dispersion of the LCP material, one can calculate the standard deviation of the retardance over the filter fractional bandpass, $\epsilon$. An estimate of the total attenuation $A$ (integrated over the whole field of view), can then be calculated from $\epsilon$ \citep{Mawet2009}: $A = \epsilon^2/4$. Table~1 summarizes our total attenuation predictions and the peak-to-peak attenuation measurements presented in the following Sect.~\ref{sect:lab}. Indeed, a coronagraph's ability to suppress the starlight is conveniently quantified and measured by the peak-to-peak attenuation, which is the ratio of the maximum in the post-coronagraphic stellar residual image to the maximum (on-axis) intensity in the direct stellar image. This metric is especially meaningful on sky, since its inverse directly represents the allowable gain in integration time for individual unsaturated exposures. The peak-to-peak attenuation is close to but can be slightly smaller than the total attenuation (integrated over the whole image), owing to the spreading of the residual PSF in an aberrated system.
\begin{deluxetable*}{cccccc}
\tabletypesize{\scriptsize}
\tablecaption{Attenuation predictions, plus laboratory and PHARO measurements. PHARO is the infrared cryogenic camera of the Palomar 5-m Hale telescope \citep{Hayward01}. The Br$\gamma$ filter is centered at $2.166 \mu$m, so it is equivalent to a $\sim 5\%$ filter centered at $2.225 \mu$m for our K-band device. BW refers to bandwidth. PTP $A$ refers to peak-to-peak attenuation.}
\tablewidth{\textwidth}
\tablehead{
\colhead{Filter} & \colhead{Filter manuf.~(lab/PHARO)} & \colhead{Central $\lambda$ ($\mu$m)} & \colhead{BW} & \colhead{Predicted total $A$} & \colhead{Measured PTP $A$} }
\startdata
H 	&JDSU (lab) &1.65 	&12\%	&$\sim 4\times 10^{-3}$ &$\sim 5\times 10^{-3}$ \\
K$_s$ &Barr(lab)/OCLI(PHARO)	&2.15 	&15\%/14\%	&$\sim 5\times 10^{-3}$ &$\sim 4\times 10^{-3}$/$\sim 3\times 10^{-3}$ \\
Br$\gamma$ &Barr (lab)	&2.166 	&1\% ($\sim 5\%$ equivalent)	&$\sim 1\times 10^{-3}$ &$\sim 2.5\times 10^{-4}$  \enddata
\end{deluxetable*}

\section{Lab setup and results}\label{sect:lab}

We tested our new devices at JPL, on our transmissive infrared coronagraphic testbench \citep{Mawet2009}. To simulate starlight, we used a supercontinuum white-light source (Fianium SC450) coupled to standard telecom monomode fibers (SMF-28e) to provide a clean spatially filtered wavefront as input to the coronagraph. The testbench uses standard AR-coated achromatic near-IR lenses with low frequency errors ($<\lambda/10$ PTV over $10\, mm$ in diameter). The entrance pupil is defined by an iris diaphragm. To filter pupil (Lyot) plane, we used another diaphragm with a diameter set to be smaller than the entrance pupil by about $10\%$ (providing an optical throughput of $\sim 81\%$). All images were recorded with a LN$_2$ cooled 12-bit Merlin InSb camera. The cold filter in the camera is a Spectrogon lowpass filter (cut-off $\sim 2.6$ $\mu$m). We oversampled the images relative to $F \lambda$ to allow a proper peak-to-peak attenuation measurement. We adopted an $F$ number of 40, similar to that planned for future telescope instruments such as SPHERE \citep{Beuzit2007}. We first built up a coronagraphic image with the light source image centered on the coronagraphic spot by adding 100 individual exposure of 16 ms. The number of frames and the individual exposure time are chosen to balance signal with background and read-out noise. The next step was to record direct non-coronagraphic images (the mask was translated off the focal plane spot) in the same conditions. All images had a median dark frame subtracted from them. 
\begin{figure}[!t]
  \centering
 \includegraphics[width=9cm]{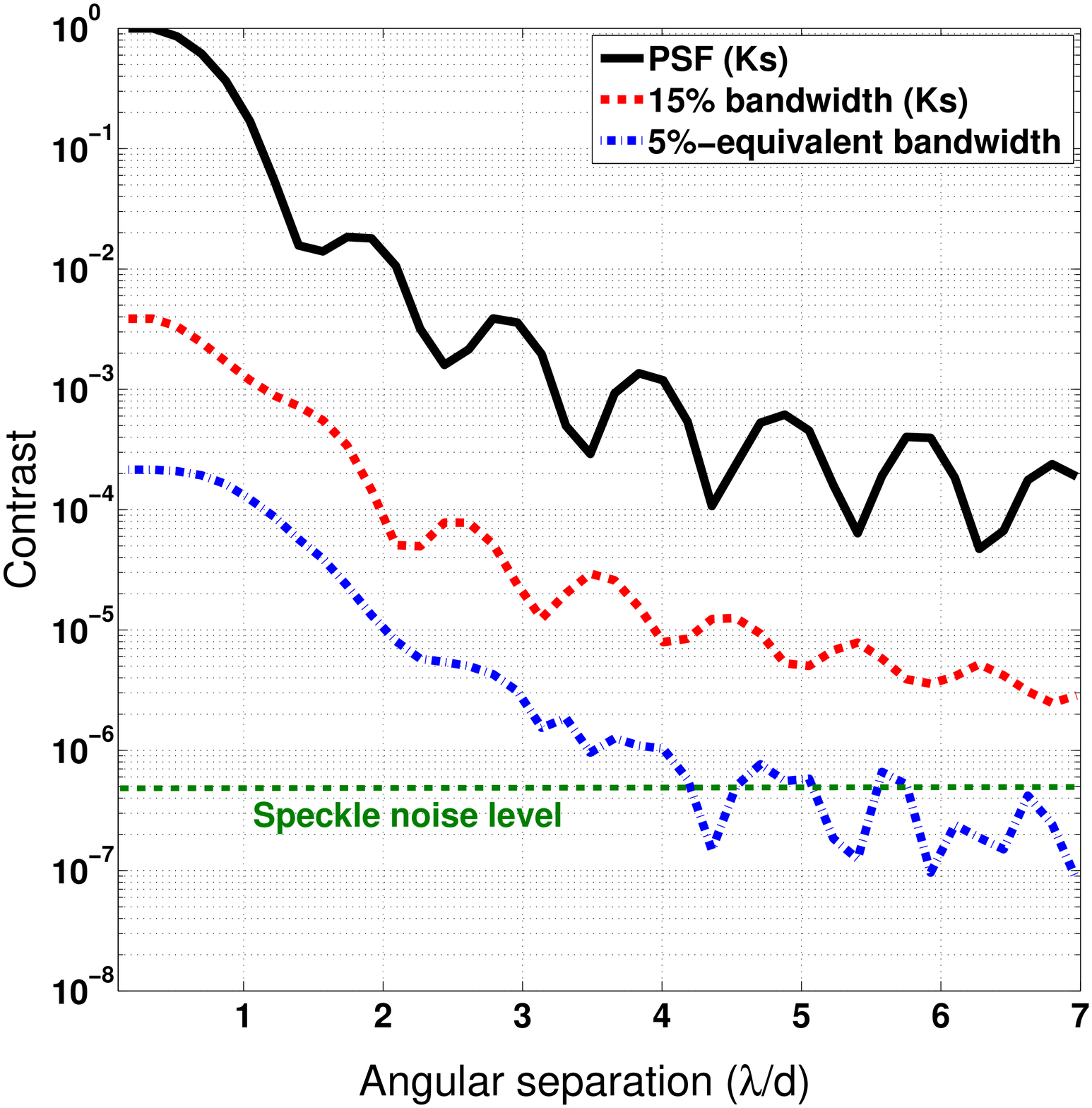}  \caption{Coronagraphic profiles measured at K$_s$ (15\% bandwidth), and in the narrowband $Br\gamma$ filter centered at $2.166$ $\mu$m, equivalent to a 5\%-wide filter centered at $\sim 2.225 \mu$m, the measured central wavelength of the K mask. The peak-to-peak attenuation is of $\sim 4\times 10^{-3}$ in the 15\% bandwidth and $\sim 2.5\times 10^{-4}$ in the reduced 5\% bandwidth. The contrast at the 3 $\lambda/d$ angular separation is $\sim 3 \times 10^{-5}$ for a 15\% bandwidth and $\sim 2 \times 10^{-6}$ for a 5\% bandwidth. Our speckle noise level on the bench is $\sim 5\times 10^{-7}$.\label{infrared_res}}
\end{figure}

Results in terms of azimuthally averaged coronagraphic profiles (radial cut in the non-attenuated and attenuated point spread functions) are shown for the K-band mask in Fig.~\ref{infrared_res}. The peak-to-peak attenuation is $\sim 4\times 10^{-3}$ in the 15\% filter and $\sim 2.5\times10^{-4}$ in the reduced 5\%-equivalent filter. The contrast near 3 $\lambda/d$ angular separation is $3 \times 10^{-5} \pm 0.05 \times 10^{-5}$ for 15\% and $2 \times 10^{-6} \pm 0.5 \times 10^{-6}$ for  5\% bandwidth. In the H band, we obtained similar results ($\sim 5\times 10^{-3}$ peak-to-peak attenuation). Our speckle noise level on the bench is $\sim 5\times 10^{-7}$, which can be considered as the error bar on all our contrast measurements. Our laboratory results are consistent with our model of the device, taking into account the residual chromaticity of the LCP material (Sect.~\ref{sec:tech}), as well as the diffraction effect of the Aluminum spot, which is negligible here (Table~1). These excellent results demonstrate that these devices are ready for use on a telescope, where the atmosphere will likely be the main limitation.

\section{On-sky results}\label{sect:onsky}

We next installed our K-band VVC mask in the cryogenic infrared camera PHARO \citep{Hayward01}. The first tests on the 5-m Hale telescope were performed with the WCS on 30 May 2009. The Palomar WCS has been described earlier \citep{Serabyn07}. In short, an unvignetted, off-axis WCS is provided by means of a set of relay optics upstream of the Palomar AO system (PALAO). The off-axis relay magnifies and shifts an off-axis sub-aperture pupil onto the deformable mirror, yielding roughly 10 cm actuator spacing in the selected pupil. This configuration reduces the effective aperture size from 5~m to 1.5~m, but allows operation in the ExAO regime. While providing unprecedented image quality on a ground-based telescope (Strehl ratios, $S\geq 90\%$), this off-axis setup is also ideal for use with phase-mask coronagraphs because these coronagraphs work best with unobscured apertures, and their very small IWA partly compensates for the loss in resolution due to the smaller aperture. 
\begin{figure*}[!t]
  \centering
\includegraphics[height=7cm]{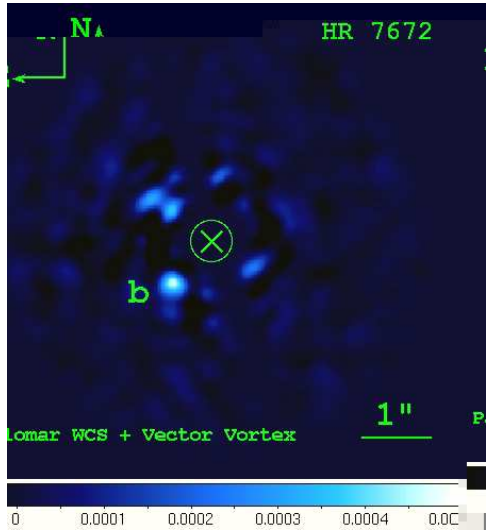}\hspace{1.5cm}\includegraphics[height=7.5cm]{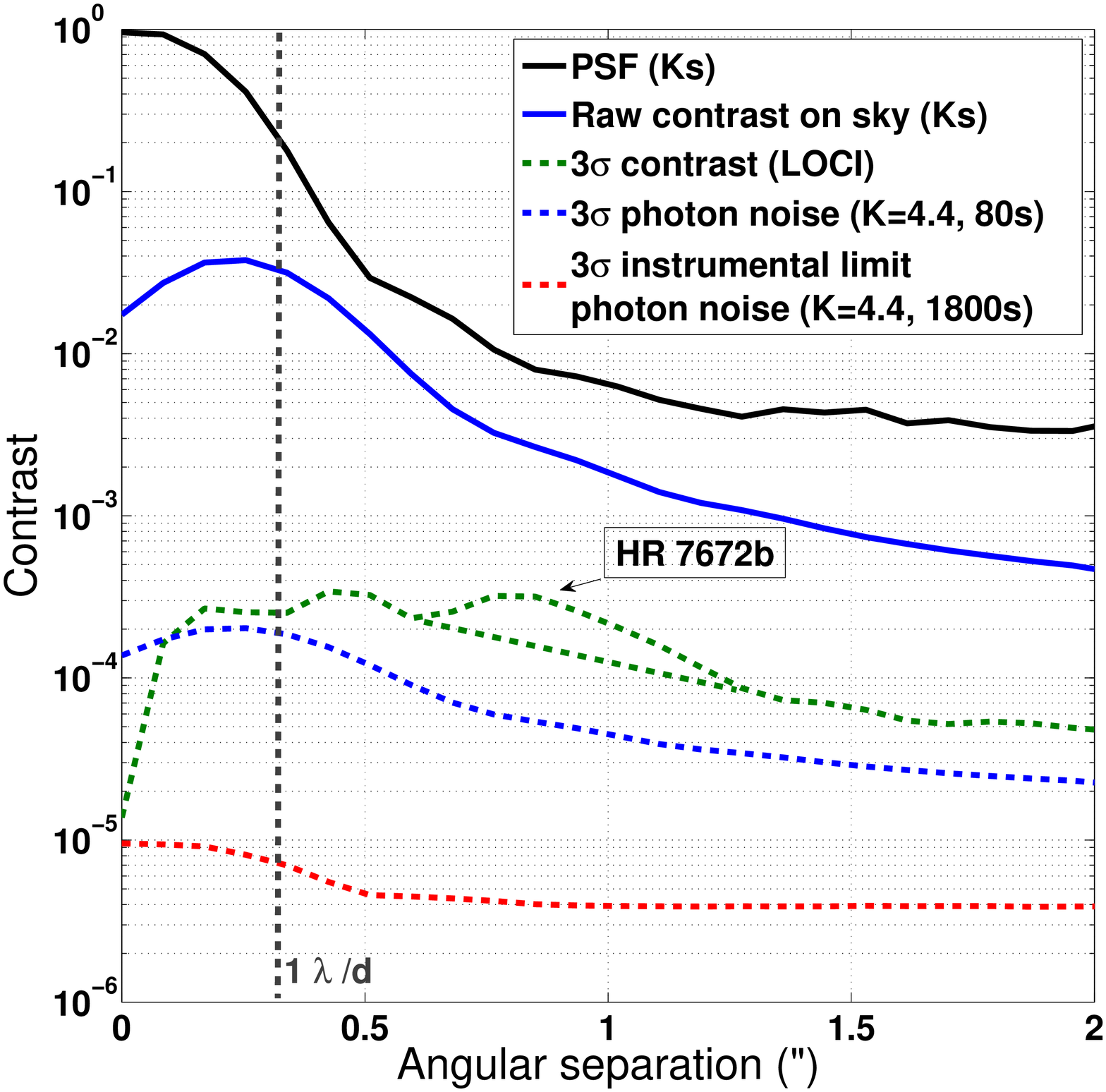} 
  \caption{Left: Detection of the brown dwarf around HR~7672 in the K$_s$ band with the VVC and the WCS at Palomar observatory. The companion lies at only $\sim 2.5 \lambda/d$, and the contrast ratio is $\sim 3\times 10^{-4}$. The stretch is linear and the scale is calibrated in contrast. Right: contrast curves of the VVC on sky. The blue curve is the raw contrast obtained on sky ($\sim 4\times 10^{-2}$ peak-to-peak attenuation). Its associated photon noise (in 80s) is shown in the bottom dotted blue curve. The intermediate green dashed curve shows the speckle noise associated with the LOCI image. The black continuous curve show a radial cut in the K$_s$-band PSF. The dashed red curve shows the photon noise limit (K=4.4, 1800s) extrapolated from the measured internal AO white light source contrast. \label{contrast}}
\end{figure*}

We first conducted contrast measurements with PALAO and PHARO attached to the Cassegrain focus of the 5-m Hale telescope, and with PALAO locked on the internal white light source. The Strehl ratio of this instrumental configuration is $S\approx 99\%$. Using PHARO's 14\%-wide K$_s$ filter, we measured a peak-to-peak attenuation $\sim 3\times 10^{-3}$, comparable to our lab measurement (Fig.~\ref{infrared_res}, Table~1). 
 
We next pointed the telescope at the star HR~7672 (K = 4.4, V = 5.8), which possesses a brown dwarf companion \citep{Liu02}. The companion has a small angular offset and high contrast ratio, r = 790 $\pm$ 5 mas, and $\Delta$K = 8.62 $\pm$ 0.07 mag ($\sim 3\times 10^{-4}$), making it the ideal test case target. We took a series of twenty 4-second exposures with the VVC centered on the target to within a few mas. We immediately followed this sequence by the same series of exposures on a reference star, with the same V-K color and brightness as the target. Matching V magnitudes is important to ensure a similar AO corrections. K magnitudes also need to be matched to ensure a proper signal-to-noise ratio match for the quasi-static speckle subtraction. Also, to reduce orientation and flexure-induced variations as much as possible, the calibration star was chosen and observed at the same hour angle and elevation as the target star. 

Fig.~\ref{contrast} (left) shows our image of HR~7672 after post-processing with the locally optimized combination of images algorithm \citep[LOCI,][]{Lafreniere07}, which is one of the most effective speckle subtraction techniques. The vortex coronagraph, owing to its absence of features anywhere in the field of view allowed us to make an even cleaner detection (in only 80s) than the four-quadrant phase-mask \citep{Serabyn09}, even in the presence of poor seeing. Note that for the WCS, the companion lies at only $\sim 2.5 \lambda/d$.

Fig.~\ref{contrast} (right) shows the radial cut in the unattenuated PSF (black), and contrast curves, starting with the raw on-sky contrast (continuous blue curve), and its associated 3-$\sigma$ photon noise for an 80s-integration time (dashed blue line). The dashed green curve represents the detection level of the final image after LOCI processing, which effectively brings us very close to the photon noise limit.

Note that the peak-to-peak attenuation ($\sim 4\times 10^{-2}$) is an order of magnitude worse than the instrumental limit measured on the internal AO source ($\sim 3\times 10^{-3}$). This difference can be explained by three major observational limitations: 
\begin{itemize}
\item [-] The seeing was an average of $1\farcs 3$, and $S$ was $\sim 85\%$. $S$ allows us to roughly estimate the attenuation by substituting the residual wavefront phase variance $\sigma^2$ from Marechal's approximation $S = e^{-\sigma^2}$, in the attenuation formula of Sect.~\ref{sec:tech} \citep{Boccaletti04}, yielding $A=-\ln{S}/4 \approx 4\times 10^{-2}$, which is consistent with the measured peak-to-peak attenuation.
\item [-] Residual pointing offsets affected the consistency of the centering of the star on the mask.
\item [-] On the sky, the vortex was slightly out of focus.
\end{itemize}
The first limitation is contingent upon the atmospheric turbulence conditions, i.e. the seeing and the turbulence coherence time. The last two are instrumental systematics that can be corrected to yield an improvement in raw contrast. Our $3\sigma$ projected instrumental contrast limit (for a K=4.4 magnitude star in 1800s) extrapolated from the white light source contrast measurement is shown in Fig.~\ref{contrast} (right, dashed red curve). This limit can be reached in practice under more favorable seeing conditions ($<1''$).

\section{Conclusions}\label{sect:concl}

We have demonstrated that a vector vortex is an extremely powerful and yet simple coronagraph. The discovery space of the VVC is maximal, with a small IWA ($\sim 1.5 \lambda/d$), and a theoretical optical throughput larger than 90\%. Thanks to ground-breaking advances in LCP technology, a set of usable high performance VVC devices was produced in a very short time. We have demonstrated laboratory contrasts on the order of a few $10^{-5}$ at an angular separation of $\sim 3\lambda/d$ over $15\%$ bandwidth. Our best result in a $5\%$-equivalent bandwidth is $\sim 2\times 10^{-6}$ at $3\lambda/d$. On sky, we easily detected a known brown dwarf at $\sim 2.5 \lambda/d$ (for the WCS), with a contrast ratio of $\sim 3\times 10^{-4}$. Our current generation of near-infrared devices is in fact already compliant with the specifications of current and next-generation ground-based instruments for exoplanet imaging and characterization such as SPHERE \citep{Beuzit2007}, GPI \citep{Macintosh2007}, Palm-3000 \citep{Dekany2006}, and HiCIAO \citep{Tamura2006}. 

The LCP technology also applies at the optical wavelengths, and we have been pursuing that route as well. The tests of our initial visible device showed good contrast results on the High Contrast Imaging Testbed \citep[HCIT,][]{Trauger07}. Using wavefront control and a topological charge 4 VVC centered at 800 nm, a dark hole with $\sim10^{-7}$ raw contrast was created in a $10\%$ bandpass \citep{Mawet09b}. Our further development of improved visible devices is ultimately aimed at providing a close to ideal coronagraph for exoplanet space missions such as the Actively Corrected Coronagraph for Exoplanet Space Studies \citep[ACCESS,][]{Trauger08}, paving the way towards the Terrestrial Planet Finder Coronagraph \citep[TPF-C,][]{Traub06}.

\acknowledgments

This work was carried out at the Jet Propulsion Laboratory (JPL), California
Institute of Technology (Caltech), under contract with NASA. D.~M.\  is supported by an appointment to the NASA Postdoctoral Program at JPL, Caltech, administered by Oak Ridge Associated Universities through a contract with NASA. The data presented in this paper are based
on observations obtained at the Hale Telescope, Palomar Observatory, as
part of a continuing collaboration between Caltech, NASA/JPL, and Cornell University.






\end{document}